%% file: main.tex
\documentclass{ifacconf}

\usepackage{graphicx} 
\usepackage{natbib}   
\usepackage{amssymb}  
\usepackage{amsmath}  
\usepackage{algorithm}
\usepackage{algpseudocode}
\usepackage{nicematrix}
\usepackage{subcaption}
\usepackage{booktabs}
\usepackage{multirow}
\usepackage{bm}

\renewcommand{\captionwidth}{\linewidth}

\definecolor{DodgerBlue}{RGB}{30,144,255}
\definecolor{MediumOrchid}{RGB}{186,85,211}
\definecolor{RoyalPurple}{RGB}{120,81,169}
\definecolor{DarkOrange}{RGB}{255,140,0}

\makeatletter
\let\IFAC@orig@ssect\@ssect
\makeatother

\usepackage{hyperref}
\hypersetup{
  colorlinks=true,
  linkcolor=red,
  citecolor=blue,
  urlcolor=blue
}

\makeatletter
\def\ps@copyright{%
  \let\@mkboth\@gobbletwo
  \def\@oddhead{\@logohead}%
  \let\@evenhead\@oddhead
  \def\@oddfoot{%
    \raisebox{-8pt}[0pt][0pt]{%
      \parbox{\textwidth}{%
        \centering\normalfont\fontsize{10}{10}\selectfont
        \textcopyright{} 2026 the authors. This work has been accepted to IFAC for publication under a Creative Commons Licence CC-BY-NC-ND.}}}%
  \let\@evenfoot\@oddfoot
}
\makeatother

\makeatletter
\def\@ssect#1#2#3#4#5#6{%
  \NR@gettitle{#6}%
  \IFAC@orig@ssect{#1}{#2}{#3}{#4}{#5}{#6}%
}
\makeatother

\makeatletter
\algnewcommand{\LineComment}[1]{%
  \Statex \hskip\ALG@thistlm {\footnotesize\color{gray}\(\triangleright\)\ \textit{#1}}%
}
\makeatother
\makeatletter
\algnewcommand{\LineCommentFirst}[1]{%
  \Statex \hskip\dimexpr\ALG@thistlm+\algorithmicindent\relax
  {\footnotesize\color{gray}\(\triangleright\)\ \textit{#1}}%
}
\makeatother

\algrenewcommand\algorithmiccomment[1]{%
  \unskip\hspace{0.5em}{\footnotesize\color{gray}\(\triangleright\)\ \textit{#1}}%
}

\begin{document}
\begin{frontmatter}

\title{Moving Horizon Estimation for Underwater Target Tracking Based on Time-Difference-of-Arrival Measurements \thanksref{footnoteinfo}} 

\thanks[footnoteinfo]{This work was supported 
                      by FCT funding PRT/BD/155064/2023 (DOI: 10.54499/PRT/BD/155064/2023),
                      LARSyS FCT funding (UID/50009/2025: DOI 10.54499/UID/50009/2025; LA/P/0083/2020: DOI 10.54499/LA/P/0083/2020),
                      and the FCT India-Portugal project RAIECO (ID: DRI/India/0699/2020).}

\author[First]{Anton Tolstonogov}
\author[First]{David Cabecinhas}
\author[First]{Pedro Batista}
\author[First]{Antonio Pascoal}

\address[First]{Institute for Systems and Robotics (ISR), LARSyS, Instituto Superior Técnico (IST),
                University of Lisbon, Portugal \\
                (e-mail: anton.tolstonogov@tecnico.ulisboa.pt, dcabecinhas@isr.tecnico.ulisboa.pt, pbatista@isr.tecnico.ulisboa.pt, antonio.pascoal@tecnico.ulisboa.pt)}

\begin{abstract}
There has been a flurry of activity in the development of robotic systems to localize and track underwater man-made or natural targets based on sparse acoustic data.
Compelling examples include the development of surface tracking systems to aid in the navigation of groups of underwater vehicles performing environmental monitoring missions or to study the motion patterns of large underwater fauna.
With current technology, the latter case can only be tackled using Time-Difference-of-Arrival (TDoA) techniques.
Recent progress in nonlinear state estimation indicates that optimization-based methods may overcome the limitations of classical recursive filtering.
However, achieving reliable estimator performance in the case of nonlinear target dynamics and sparse measurements remains a key challenge.
In this paper, we study a Moving Horizon Estimation (MHE) approach to TDoA-based underwater target tracking.
Through a 2D simulation environment capturing typical marine conditions, we show that the MHE-based estimator maintains reliable tracking in the considered scenarios even when the classical EKF becomes unreliable.
The results highlight that multi-step trajectory coupling and physically consistent constraints, which are key advantages of the MHE approach, significantly enhance estimator robustness.
It is shown that the MHE approach offers promise as a practical and scalable building block for future multi-agent tracking systems based on TDoA measurements operating in real underwater missions.
\end{abstract}


\end{frontmatter}

\input{part1_introduction.tex}

\input{part2_problem.tex}

\input{part3_estimator_mhe.tex}

\input{part4_simulation.tex}

\input{part5_conclusion.tex}

\section*{DECLARATION OF GENERATIVE AI AND AI-ASSISTED TECHNOLOGIES IN THE WRITING PROCESS}
During the preparation of this work the authors used ChatGPT-5 in order to proofread the text.
After using this service, the authors reviewed and edited the content as needed and take full responsibility for the content of the publication.

\bibliography{references}

\end{document}

%% file: part1_introduction.tex
\section{Introduction}

There is growing interest in advanced tools and methodologies for ocean exploration, particularly for monitoring marine life and understanding how marine-species movement relates to environmental conditions.
These monitoring capabilities are essential for translating marine animal tracking data into effective conservation and resource-management decisions \cite{haysTranslatingMarineAnimal2019}, but this remains challenging due to the scale of the ocean and the limited endurance, sensing capacity, and robustness of single-robot platforms.

To address these limitations, current trends in marine robotics emphasize heterogeneous teams of cooperative marine vehicles.
Cooperative multi-vehicle systems are important for detecting, localizing, and tracking underwater fauna equipped with low-power acoustic tags, using receivers mounted on autonomous surface vehicles (ASVs), autonomous underwater vehicles (AUVs), or gliders.

Beyond wildlife monitoring, similar cooperative localization strategies are employed to provide external navigational references for survey-class AUVs, support seabed-mapping operations, assist remotely operated vehicles (ROVs), and enable adaptive ocean-sampling campaigns.
In all these scenarios, accurate underwater localization and tracking based on sparse acoustic signals is a fundamental requirement.

A widely used sensing modality in these applications is Time-Difference-of-Arrival (TDoA) localization.
Unlike range-based methods, TDoA does not require knowledge of the absolute emission time, which makes it compatible with low-power wildlife tags emitting simple, non-coded pings.
Its minimal hardware requirements and relaxed synchronization constraints also make TDoA practical for multi-vehicle deployments.

\noindent\textbf{Related work:} TDoA-based localization has been successfully applied across a wide range of underwater and terrestrial domains, including marine animal tracking \cite{espinozaTestingNewAcoustic2011}, distributed target localization \cite{ennasrTimeDifferenceofArrivalTDOABasedDistributed2020}, cooperative AUV surveying, docking \cite{batistaTimeDifferencesArrivalbased2010}, wireless ranging radar, and cellular positioning \cite{dong-hoshinComparisonsErrorCharacteristics2002}.

Despite these advantages, TDoA-based target localization remains challenging because the measurement model is highly nonlinear and must be combined with nonlinear and often nonholonomic motion models of either the tracking vehicles or the tagged animals.
In dynamic scenarios, state-estimation methods such as the Extended Kalman Filter (EKF) \cite{sriramTDoABasedEKF2016}, the Unscented Kalman Filter, and particle-filter approaches \cite{bordoyRobustTrackingMobile2013} are commonly used.
However, filtering techniques relying on local linearization, particularly the EKF, often degrade when the system operates far from the linearization point, when measurements are sparse, or when the TDoA receiver geometry yields poor observability.

These limitations motivate optimization-based estimation frameworks, particularly Moving Horizon Estimation (MHE).
MHE naturally incorporates nonlinear models, state and input constraints, and multi-step observational windows.
Prior work \cite{raoConstrainedStateEstimation2003,haseltineCriticalEvaluationExtended2005} has demonstrated that MHE can outperform EKF-based methods in nonlinear settings and under low-rate or noisy measurements.
MHE has also shown promise in cooperative underwater navigation using acoustic range measurements \cite{wangCooperativeLocalizationAUVs2014}.
However, MHE has not yet been systematically studied for underwater target tracking using TDoA measurements with nonlinear motion models representative of tagged animals or marine vehicles.

\noindent\textbf{Contribution:} This paper formulates and evaluates an MHE approach for underwater target tracking using TDoA measurements and compares it with a classical EKF baseline under a number of operational conditions.

The main contributions of this paper are:
\begin{itemize}
    \item a complete formulation of the nonlinear MHE problem for TDoA-based target tracking with nonholonomic motion models;
    \item a quantitative comparison of MHE and EKF performance across diverse conditions, including target motion patterns, initialization errors, and measurement sparsity.
\end{itemize}

The remainder of the paper is organized as follows.
Section~\ref{sec:problem} presents the problem formulation and the TDoA measurement model.
Section~\ref{sec:mhe} details the centralized MHE formulation for TDoA tracking.
Section~\ref{sec:simulation} reports numerical experiments comparing the proposed MHE with an EKF baseline across multiple scenarios.
Section~\ref{sec:conclusion} concludes the paper and outlines future work.

%% file: part2_problem.tex
\section{Problem Formulation}\label{sec:problem}

\subsection{System Description}
Consider a group of $M$ trackers tasked with tracking a moving target.
In this scenario, we assume that the target follows a simplified nonholonomic motion model with constant linear and angular velocities.
The target state is described by its planar position $\bm{p} = [x,\,y]^\top$ and heading $\psi$ in the inertial frame $\{\mathcal{I}\}$, together with its linear velocity $v$ and angular velocity $r$ expressed in the body frame $\{\mathcal{B}\}$ attached to the target.

\subsubsection{Target Model}
Using Euler-based discretization, we model the target dynamics in the standard discrete-time nonlinear state-space form with additive noise as
\begin{equation}
   \label{eq:state_space_model}
   \begin{aligned}
   \bm{x}_{k+1} &= f(\bm{x}_k) + \bm{w}_k, \\[2pt]
   \bm{y}_k &= h(\bm{x}_k) + \bm{\nu}_k,
   \end{aligned}
\end{equation}
\vspace{-1em}
\noindent where
\begin{itemize}
   \item $\bm{x}_k = [\bm{p}_k^\top,\, \psi_k,\, v_k,\, r_k]^\top \in \mathbb{R}^5$ is the target state vector at time step $k$,
   \item $\bm{p}_k \in \mathbb{R}^2$ is the target position in the inertial frame at time step $k$,
   \item $f(\bm{x}_k)$ is the nonlinear kinematic state-transition model,  
   \item $\bm{w}_k \sim \mathcal{N}(\bm{0},\, Q)$ is Gaussian process noise with zero mean and covariance matrix $Q \in \mathbb{R}^{5\times5} \succ 0$,
   \item $\bm{y}_k \in \mathbb{R}^{(M-1)}$ is the vector of TDoA measurements,
   \item $h(\bm{x}_k)$ is the nonlinear TDoA measurement function,
   \item $\bm{\nu}_k \sim \mathcal{N}(\bm{0},\, R)$ is Gaussian measurement noise with covariance matrix $R \in \mathbb{R}^{(M-1)\times (M-1)} \succ 0$.
\end{itemize}

Both noise terms $\bm{w}_k$ and $\bm{\nu}_k$ are assumed mutually independent and temporally uncorrelated.
The discrete-time kinematics are given by
\begin{equation}
   \label{eq:motion_model}
   f(\bm{x}_k) =
   \begin{bmatrix}
   \bm{p}_k + \Delta T\, v_k \bm{e}(\psi_k) \\
   \psi_k + \Delta T\, r_k \\
   v_k \\
   r_k
   \end{bmatrix},
\end{equation}
\noindent where $\Delta T$ is the discretization step and
$
\bm{e}(\psi_k) \triangleq
\begin{bmatrix}
\cos\psi_k & \sin\psi_k
\end{bmatrix}^\top
$
is the unit vector aligned with the target heading in the inertial frame.

\subsubsection{Measurement Model}
According to TDoA measurement theory \cite{sathyanAlgorithmsPassiveLocalization2007}, the measurement function $h(\bm{x}_k)$ maps the target position to TDoA observations for a group of trackers, that is,
\begin{equation}
   \label{eq:tdoa}
   h(\bm{x}_k)
   =
   \begin{bmatrix}
   h_1(\bm{x}_k) & \cdots & h_{M-1}(\bm{x}_k)
   \end{bmatrix}^\top .
\end{equation}
For $i = 1,\ldots,M-1$, each scalar TDoA component is defined as
\begin{equation}
   \label{eq:tdoa_component}
   h_i(\bm{x}_k)
   =
   \frac{1}{v_s}
   \left(
   \|\bm{p}_k - \bm{p}_k^{[1]}\|
   -
   \|\bm{p}_k - \bm{p}_k^{[i+1]}\|
   \right),
\end{equation}
where $v_s$ is the known speed of sound, $\bm{p}_k^{[i]}$ for $i \in \{1,\ldots,M\}$ is the position of tracker $i$, and $\bm{p}_k^{[1]}$ denotes the reference tracker.
The latter is the tracker with respect to which the TDoA measurements are computed.
It can be chosen arbitrarily, but for simplicity, we use tracker $1$ as the reference in \eqref{eq:tdoa}.

%% file: part3_estimator_mhe.tex
\section{MHE Formulation}\label{sec:mhe}

Following the MHE framework proposed in \cite{schillerLyapunovFunctionRobust2023}, the target-tracking problem at time $k$ is posed as a finite-horizon time-discounted optimization problem over a recent state trajectory.
Let $N$ denote the measurement horizon, so that the active window contains state estimates at $N+1$ time instants and $N$ model transitions, and define the beginning of the active window as
\[
k_0 \triangleq k-N .
\]
For notational simplicity, the full-horizon case is presented; during start-up, the effective horizon length can be replaced by $\min\{k,N\}$.
The state trajectory over this window is parameterized by
\[
\bm{\chi}_{i|k} \triangleq \hat{\bm{x}}_{k_0+i|k},
\qquad i=0,\ldots,N,
\]
where $\bm{\chi}_{i|k}$ denotes the estimate of the target state at time instant $k_0+i$ after processing measurements through time $k$.
The optimized trajectory of estimated states over the active window is written compactly as
\[
\bm{\chi}_{\cdot|k}
\triangleq
(\bm{\chi}_{0|k},\bm{\chi}_{1|k},\ldots,\bm{\chi}_{N|k}).
\]

At each time step, the estimator seeks the state trajectory that best explains the observed TDoA measurements while remaining compatible with the discrete-time motion model.
This leads to a discounted multiple-shooting cost composed of three terms: (i) a measurement term penalizing discrepancies between predicted and observed TDoA values, (ii) a kinematic term penalizing one-step model residuals between consecutive time instants, and (iii) an arrival term summarizing information prior to the current window.
The time-discounting factor $\eta \in (0,1)$ assigns larger weight to more recent data.
Throughout this section, $\|\bm{z}\|_A^2 \triangleq \bm{z}^\top A\bm{z}$.

\subsubsection{Measurement Term}
Using the TDoA measurement model \eqref{eq:tdoa}, the measurement residual at stage $i$ is defined as
\begin{equation}
\label{eq:mhe_measurement_residual}
\bm{e}^y_i
\triangleq
h(\bm{\chi}_{i|k}) - \bm{y}_{k_0+i},
\qquad i=0,\ldots,N,
\end{equation}
where the dependence of $h(\cdot)$ on the known tracker positions at time $k_0+i$ is omitted for compactness.
The corresponding discounted measurement cost is
\begin{equation}
\label{eq:mhe_measurement_cost}
J_{\text{measure}}
=
\sum_{i=0}^{N}
\eta^{N-i}
\left\|\bm{e}^y_i\right\|_{R^{-1}}^2 .
\end{equation}
Thus, the most recent measurement $\bm{y}_k$ has unit weight, while older measurements are exponentially discounted.
The terminal residual at $i=N$ is included because the estimator reports the filtering estimate after receiving $\bm{y}_k$.

\subsubsection{Kinematic Term}
The kinematic term encourages consecutive trajectory states to follow the one-step prediction given by the discretized motion model \eqref{eq:motion_model}.
For each transition in the measurement horizon window, define
\begin{equation}
\label{eq:mhe_kinematic_residual}
\bm{e}^x_i
\triangleq
\bm{\chi}_{i+1|k} - f(\bm{\chi}_{i|k}),
\qquad i=0,\ldots,N-1 .
\end{equation}
The corresponding kinematic cost is given by
\begin{equation}
\label{eq:mhe_kinematic_cost}
J_{\text{kinematic}}
=
2\sum_{i=0}^{N-1}
\eta^{N-1-i}
\left\|\bm{e}^x_i\right\|_{Q^{-1}}^2 .
\end{equation}

\subsubsection{Arrival Term}
The arrival term incorporates information prior to the current horizon.
Let $\bar{\bm{x}}_{k_0}$ denote the stored filtering prior at the first time instant of the active window, summarizing measurements before time $k_0$.
Following \cite{alessandriMovinghorizonStateEstimation2008}, the arrival residual in the simplified constant-weight form is given by
\begin{equation}
\label{eq:mhe_arrival_residual}
\bm{e}^a_k
\triangleq
\bm{\chi}_{0|k} - \bar{\bm{x}}_{k_0},
\end{equation}
and the corresponding discounted arrival cost is
\begin{equation}
\label{eq:mhe_arrival_cost}
J_{\text{arrival}}
=
2\eta^N
\left\|\bm{e}^a_k\right\|_{P^{-1}}^2 .
\end{equation}
The matrix $P \succ 0$ represents the arrival-prior covariance, so $P^{-1}$ weights the arrival residual in the MHE cost.
Together with $Q$ and $R$, it defines the relative uncertainty assigned to prior information, model consistency, and TDoA measurements.

Combining the three terms yields the trajectory cost
\begin{equation}
\label{eq:mhe_discounted_cost}
J_k(\bm{\chi}_{\cdot|k})
=
J_{\text{arrival}}
+
J_{\text{kinematic}}
+
J_{\text{measure}} .
\end{equation}
The complete MHE problem is formulated as
\begin{equation}
\label{eq:mhe_problem}
\begin{aligned}
\bm{\chi}_{\cdot|k}^\star
=
\arg\min_{\bm{\chi}_{\cdot|k}} \ &
J_k(\bm{\chi}_{\cdot|k}) \\
\text{s.t.} \ &
\bm{\chi}_{i|k} \in \mathcal{X},
\  i=0,\ldots,N .
\end{aligned}
\end{equation}
The feasible set $\mathcal{X}$ denotes the bounded admissible state set.
It encodes the operational area, physical constraints, and prior knowledge about the target motion.

This work uses the MHE in a filtering form.
After receiving $\bm{y}_k$, the target state at time $k$ is estimated as
\[
\hat{\bm{x}}_{k|k}
\triangleq
\bm{\chi}_{N|k}^\star .
\]
Only this current-time value is used as the reported target-state estimate.
Nevertheless, the full estimated trajectory $\bm{\chi}_{\cdot|k}^\star$ is retained to initialize the next MHE problem: when the next measurement becomes available, the trajectory is propagated through the kinematic model by one time step to provide a warm start.
Because the problem is nonlinear, the numerical solver generally returns a local optimum.
In practice, warm starts from the previous horizon and bounded state constraints reduce sensitivity to poor local minima.

For the bounded-noise case, time-discounted MHE admits stability guarantees.
In particular, following \cite{schillerLyapunovFunctionRobust2023}, if the nonlinear system admits an exponential $\delta$-IOSS certificate on the admissible set, then for a sufficiently long estimation horizon the corresponding time-discounted MHE is robustly globally exponentially stable (RGES).

%% file: part4_simulation.tex
\section{Simulation}\label{sec:simulation}

In this section, we numerically evaluate the proposed MHE formulation against the EKF baseline under a variety of motion patterns, measurement conditions, and initialization errors.
We first describe the simulation setup and evaluation metrics, and then report localization accuracy and computational performance.

\subsection{Simulation Setup}\label{sec:sim_setup}

All experiments use synthetic data generated in controlled simulations.
We consider three representative motion scenarios for the target.
\textit{In the first scenario} (constant), the target moves with constant linear and angular velocities.
\textit{In the second scenario} (UUV-like\footnote{Unmanned Underwater Vehicle}), the motion also follows nominally constant linear and angular velocities but includes occasional sharp heading changes that mimic the ``lawn-mowing'' or survey-style maneuvers commonly executed by UUVs.
\textit{In the third scenario} (whale-like), we emulate the behavior of large marine mammals by allowing the linear and angular velocities to evolve as slowly varying random processes, resulting in smooth but unpredictable trajectories.

\begin{figure}[htbp]
   \centering
   \includegraphics[width=\linewidth]{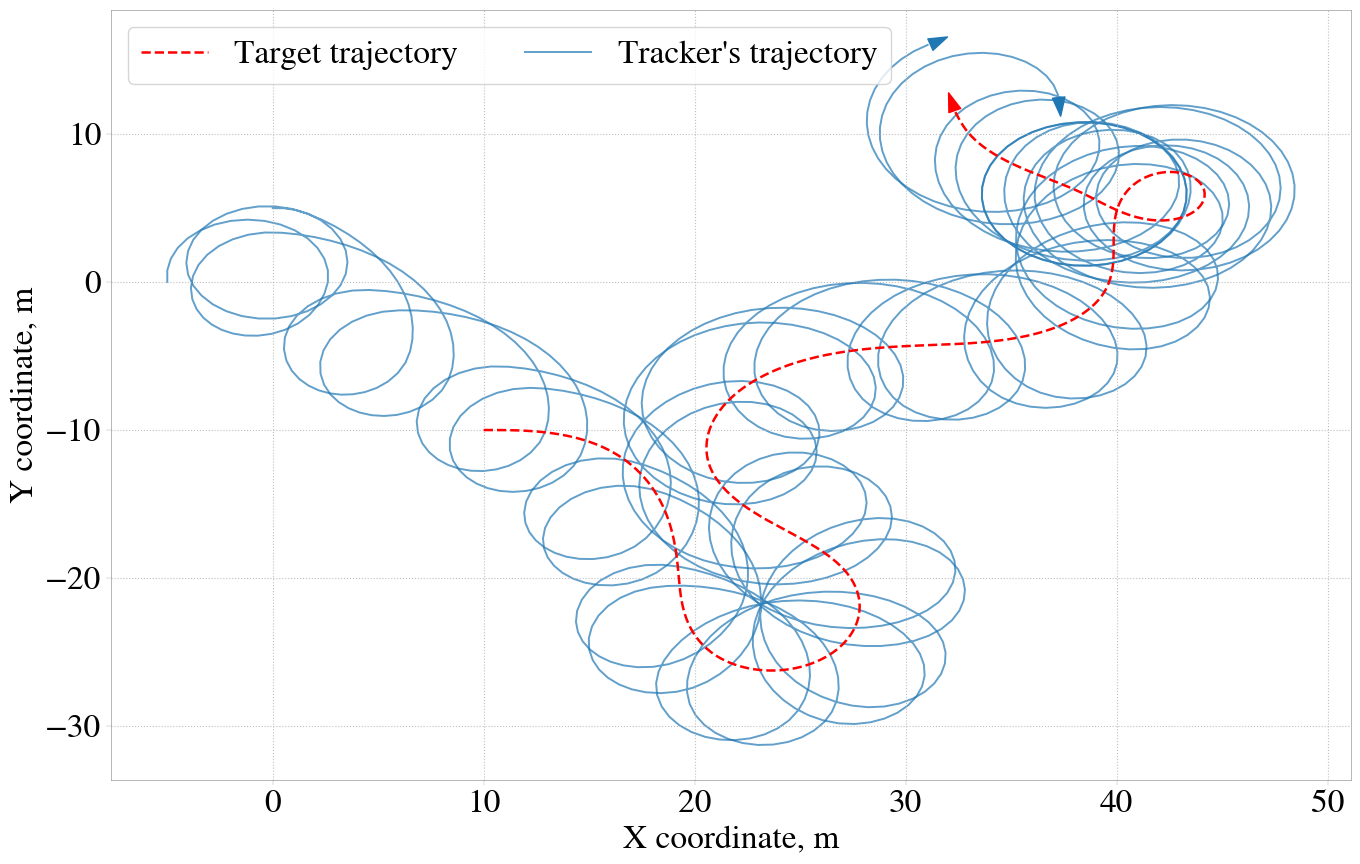}
   \caption{Example of a single target trajectory and epicyclic tracker trajectories for the whale-like scenario.} 
   \label{fig:trajectory}
\end{figure}

The trackers circulate around the target, forming an epicycle pattern over time. 
This motion continuously changes the tracker-target geometry, so that the TDoA information is collected from different directions.
This tends to make the accumulated measurement geometry more isotropic from an observability viewpoint.
Such circulating trajectories are widely used in range- and bearing-based localization to improve observability~\cite{hungRangebasedTargetLocalization2020}, which motivates their use in this work.
Example trajectories for the whale-like scenario are shown in Fig.~\ref{fig:trajectory}.

For simplicity, simulations take place in a 2D environment with two trackers.
Unless stated otherwise, all estimators share identical process and measurement noise parameters, horizon length, and solver tolerances.
All parameters are summarized in Table~\ref{tab:sim_params}.
Estimation accuracy is evaluated using the Root Mean Square Error (RMSE), computed separately for position ($\mathrm{pos}$), heading ($\psi$), linear velocity ($v$), and yaw rate ($r$).

The implementation of the proposed algorithm and the code used to conduct the experiments are available as open-source software on GitHub at \url{https://github.com/bioniwulf/playground-estimators-comparison}.

\begin{table}[htbp]
\centering
\captionsetup{width=\columnwidth,justification=centering}
\caption{Common simulation parameters.}
\label{tab:sim_params}
\setlength{\tabcolsep}{3pt}
\renewcommand{\arraystretch}{1.00}
\begin{tabular}{ll}
\toprule
\textbf{Parameter} & \textbf{Value} \\
\midrule
\multicolumn{2}{l}{\textit{Data generation}} \\
Total simulation time, $T$ [s] & 160.0 \\
Discretization step, $\Delta T$ [s] & 0.1 \\
Speed of sound, $v_s$ [m/s] & 1024.0 \\
Acoustic emission period [s] & 1.0 \\
TDoA noise std., $\sigma_r$ [s] & $1\times10^{-5}$ \\
\midrule
\multicolumn{2}{l}{\textit{Target motion parameters (constant scenario)}} \\
Linear velocity, $v$ [m/s] & 0.6 \\
Angular rate, $r$ [rad/s] & 0.01 \\
\midrule
\multicolumn{2}{l}{\textit{Initial states}} \\
\multicolumn{2}{l}{(\textit{State vector units}: $[{\rm m},{\rm m},{\rm rad},{\rm m/s},{\rm rad/s}]$)} \\
Initial target state, $\bm{x}_0$ & (10.0, -10.0, 0.0, 0.0, 0.0) \\
Near initial estimate, $\hat{\bm{x}}_0$    & (8.0, -8.0, 0.0, 0.2, 0.0) \\
Moderate initial estimate, $\hat{\bm{x}}_0$ & (0.0, 0.0, 0.0, 0.2, 0.0) \\
Distant initial estimate, $\hat{\bm{x}}_0$  & (-5.0, 5.0, 0.0, 0.2, 0.0) \\
\midrule
\multicolumn{2}{l}{\textit{MHE configuration}} \\
Process covariance, $Q$ & $\operatorname{diag}(0.5, 0.5, 0.1, 2, 2)$ \\
Arrival covariance, $P$ & $\operatorname{diag}(30.0, 30.0, 10.0, 1.0, 1.0)$ \\
Measurement covariance, $R$ & $\sigma_r^2 I_{M-1}$ \\
Measurement horizon, $N$ & 15 \\
Time-discounting factor, $\eta$ & 0.85 \\
Admissible set, $\mathcal{X}$ & $(\pm100,\pm100,\pm2\pi,[0,1],\pm1.047)$ \\
(\textit{state order as above}) & \\
\midrule
\multicolumn{2}{l}{\textit{Baseline EKF configuration}} \\
Process covariance, $Q$ & $\operatorname{diag}(0.5, 0.5, 0.1, 2, 2)$ \\
Initial covariance, $P_0$ & $\operatorname{diag}(30.0, 30.0, 10.0, 1.0, 1.0)$ \\
Measurement covariance, $R$ & $\sigma_r^2 I_{M-1}$ \\
\bottomrule
\end{tabular}
\end{table}

\subsection{Simulation Results: Localization Accuracy}

\begin{table}[htbp]
\centering
\captionsetup{width=\columnwidth,justification=centering}
\caption{RMSE for the constant scenario. The row blocks correspond to clean 1 s updates, 1 s updates with outliers, and sparse 3 s updates.}
\label{tab:results_cv}
\setlength{\tabcolsep}{3pt}
\renewcommand{\arraystretch}{1.00}
\begin{tabular}{lcccc}
\toprule
\multirow{2}{*}{\shortstack{Estimator\\configuration}} & \multicolumn{4}{c}{RMSE} \\
\cmidrule(lr){2-5}
& $\bm{p}$ [m] & $\psi$ [rad] & $v$ [m/s] & $r$ [rad/s] \\
\midrule
\multicolumn{5}{c}{1 s emission period} \\
\midrule
EKF (distant init.) & $4.15$ & $2.81$ & $1.20$ & $1.263$ \\
EKF (near init.)    & $\bm{1.12}$ & $0.17$ & $\bm{0.08}$ & $0.028$ \\
MHE (distant init.) & $3.47$ & $0.31$ & $0.14$ & $0.07$ \\
MHE (near init.)    & $1.16$ & $\bm{0.12}$ & $0.09$ & $\bm{0.020}$ \\
\midrule
\multicolumn{5}{c}{1 s emission period, with outliers} \\
\midrule
EKF (distant init.) & $4.73$ & $1.33$ & $0.86$ & $5.52$ \\
EKF (near init.)    & $\bm{1.14}$ & $0.30$ & $0.11$ & $0.06$ \\
MHE (distant init.) & $3.48$ & $0.38$ & $0.15$ & $0.08$ \\
MHE (near init.)    & $1.19$ & $\bm{0.24}$ & $\bm{0.10}$ & $\bm{0.05}$ \\
\midrule
\multicolumn{5}{c}{3 s emission period} \\
\midrule
EKF (moderate init.) & $1427.38$ & $1.80$ & $2.49$ & $3.87$ \\
EKF (near init.)     & $5093.72$ & $1.87$ & $1.82$ & $6.70$ \\
MHE (moderate init.) & $6.55$ & $1.43$ & $0.34$ & $0.49$ \\
MHE (near init.)     & $\bm{4.69}$ & $\bm{0.68}$ & $\bm{0.20}$ & $\bm{0.23}$ \\
\bottomrule
\end{tabular}
\end{table}

\begin{table}[htbp]
\centering
\captionsetup{width=\columnwidth,justification=centering}
\caption{RMSE for the UUV-like scenario. The row blocks correspond to clean 1 s updates, 1 s updates with outliers, and sparse 3 s updates.}
\label{tab:results_uuv}
\setlength{\tabcolsep}{3pt}
\renewcommand{\arraystretch}{1.00}
\begin{tabular}{lcccc}
\toprule
\multirow{2}{*}{Estimator} & \multicolumn{4}{c}{RMSE} \\
\cmidrule(lr){2-5}
& $\bm{p}$ [m] & $\psi$ [rad] & $v$ [m/s] & $r$ [rad/s] \\
\midrule
\multicolumn{5}{c}{1 s emission period} \\
\midrule
EKF (distant init.) & $811.12$ & $1.94$ & $3.28$ & $3.74$ \\
EKF (near init.)    & $\bm{0.67}$ & $0.30$ & $0.08$ & $\bm{0.15}$ \\
MHE (distant init.) & $2.78$ & $0.37$ & $0.24$ & $0.16$ \\
MHE (near init.)    & $0.77$ & $\bm{0.27}$ & $\bm{0.06}$ & $\bm{0.15}$ \\
\midrule
\multicolumn{5}{c}{1 s emission period, with outliers} \\
\midrule
EKF (distant init.) & $966.72$ & $1.75$ & $5.79$ & $6.38$ \\
EKF (near init.)    & $\bm{0.81}$ & $0.55$ & $0.17$ & $0.19$ \\
MHE (distant init.) & $2.80$ & $0.45$ & $0.25$ & $0.18$ \\
MHE (near init.)    & $0.92$ & $\bm{0.38}$ & $\bm{0.10}$ & $\bm{0.17}$ \\
\midrule
\multicolumn{5}{c}{3 s emission period} \\
\midrule
EKF (moderate init.) & $3863.06$ & $1.92$ & $0.90$ & $9.16$ \\
EKF (near init.)     & $17123.04$ & $1.96$ & $5.74$ & $8.45$ \\
MHE (moderate init.) & $5.66$ & $1.03$ & $0.32$ & $0.22$ \\
MHE (near init.)     & $\bm{2.50}$ & $\bm{0.51}$ & $\bm{0.13}$ & $\bm{0.13}$ \\
\bottomrule
\end{tabular}
\end{table}

\begin{table}[htbp]
\centering
\captionsetup{width=\columnwidth,justification=centering}
\caption{RMSE for the whale-like scenario. The row blocks correspond to clean 1 s updates, 1 s updates with outliers, and sparse 3 s updates.}
\label{tab:results_whale}
\setlength{\tabcolsep}{3pt}
\renewcommand{\arraystretch}{1.00}
\begin{tabular}{lcccc}
\toprule
\multirow{2}{*}{Estimator} & \multicolumn{4}{c}{RMSE} \\
\cmidrule(lr){2-5}
& $\bm{p}$ [m] & $\psi$ [rad] & $v$ [m/s] & $r$ [rad/s] \\
\midrule
\multicolumn{5}{c}{1 s emission period} \\
\midrule
EKF (moderate init.) & $187.64$ & $1.78$ & $2.08$ & $7.20$ \\
EKF (near init.)     & $1.12$ & $0.58$ & $0.29$ & $0.23$ \\
MHE (moderate init.) & $2.95$ & $0.31$ & $0.17$ & $0.09$ \\
MHE (near init.)     & $\bm{1.04}$ & $\bm{0.21}$ & $\bm{0.10}$ & $\bm{0.07}$ \\
\midrule
\multicolumn{5}{c}{1 s emission period, with outliers} \\
\midrule
EKF (moderate init.) & $1466.11$ & $1.73$ & $8.22$ & $5.20$ \\
EKF (near init.)     & $\bm{1.10}$ & $\bm{0.61}$ & $0.17$ & $\bm{0.14}$ \\
MHE (moderate init.) & $3.00$ & $0.66$ & $0.19$ & $0.21$ \\
MHE (near init.)     & $1.17$ & $0.62$ & $\bm{0.15}$ & $0.20$ \\
\midrule
\multicolumn{5}{c}{3 s emission period} \\
\midrule
EKF (moderate init.) & $4873.81$ & $1.82$ & $3.77$ & $10.74$ \\
EKF (near init.)     & $5443.76$ & $1.76$ & $1.38$ & $2.46$ \\
MHE (moderate init.) & $6.00$ & $1.39$ & $0.41$ & $\bm{0.32}$ \\
MHE (near init.)     & $\bm{5.51}$ & $\bm{1.37}$ & $\bm{0.40}$ & $\bm{0.32}$ \\
\bottomrule
\end{tabular}
\end{table}

\begin{figure}[htbp]
   \centering
   \includegraphics[width=\linewidth]{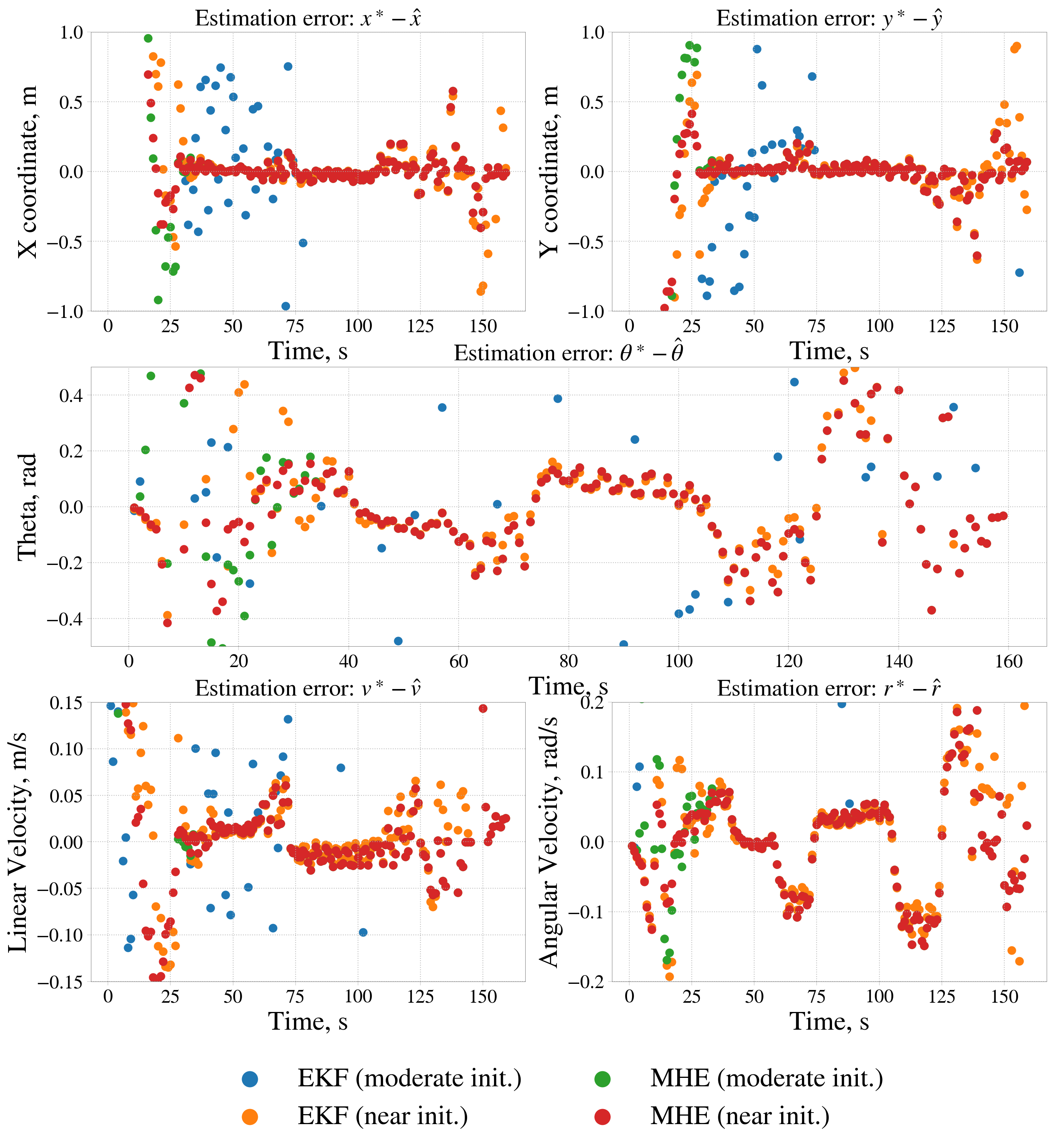}
   \caption{Example time series of position, heading, linear-velocity, and yaw-rate estimation errors for the whale-like scenario. Stars denote measurement-based updates, while dots represent kinematic model predictions.}
   \label{fig:time_series_error}
\end{figure}

Tables~\ref{tab:results_cv}--\ref{tab:results_whale} compare EKF and MHE across motion scenarios, measurement sparsity, and initialization conditions.
Figure~\ref{fig:time_series_error} shows a representative whale-like error time series.

For a $1$~s emission period without outliers and near initialization, both estimators achieve low errors.
In these favorable conditions, neither estimator dominates uniformly in the constant-velocity and UUV-like scenarios: the EKF is slightly better in position RMSE, while the MHE is better in heading and velocity-related states.
In the whale-like scenario, the MHE provides the lowest RMSE in all reported states.

The EKF is highly sensitive to initialization and often diverges under distant or moderate initial guesses, whereas the MHE remains stable in all tested cases.
Random outliers also affect the EKF more strongly under poor initialization, while the MHE maintains bounded errors because the horizon cost reduces the impact of isolated corrupted measurements.
For near initialization, the advantage under outliers is more scenario dependent: in the constant and UUV-like cases the MHE performs better in most state components, whereas in the whale-like case the two estimators become comparable and the EKF is slightly better in position, heading, and yaw-rate RMSE.

Sparse acoustic updates further separate the two estimators: increasing the emission period to $3$~s severely degrades EKF performance and often leads to divergence, while the MHE remains stable with bounded errors by using the process model and multi-step consistency over the horizon.
Across all three motion scenarios, the MHE degrades much more gracefully than the EKF as measurements become sparse or initialization becomes less accurate.

\subsection{Computational Aspects}
\label{sec:computational_aspects}

Performance is a critical factor for real-time underwater applications.
The MHE was implemented in CasADi~\cite{anderssonCasADiSoftwareFramework2019} and solved with IPOPT~\cite{wachterImplementationInteriorpointFilter2006}.
Timing tests were run on an Intel i9-14900HX CPU @~5.8\,GHz for clean measurements, a 1\,s emission period, and the near initial estimate.

Table~\ref{tab:exec_time} reports statistics averaged over three independent runs.

\begin{table}[htbp]
  \centering
  \captionsetup{width=\columnwidth,justification=centering}
  \caption{
    Execution-time statistics for the three motion scenarios.
    \textit{Cold start} denotes the wall time of the first optimization solve.
    \textit{Sliding window} reports the mean and the 90th percentile (P90) over the last $50$ warm-started updates.
  }
  \label{tab:exec_time}
  \setlength{\tabcolsep}{3pt}
  \renewcommand{\arraystretch}{1.05}
  \begin{tabular}{lcccc}
  \toprule
  \multirow{2}{*}{Scenario} 
    & \multirow{2}{*}{Cold start time [ms]} 
    & \multicolumn{2}{c}{Sliding window time [ms]} \\
  \cmidrule(lr){3-4}
    &  & mean & P90 \\
  \midrule
  Constant     & $3.24$ & $3.42$ & $5.75$ \\
  UUV-like     & $4.77$ & $4.05$ & $4.76$ \\
  Whale-like   & $3.20$ & $3.88$ & $6.81$ \\
  \bottomrule
  \end{tabular}
\end{table}

After warm-up, the MHE maintains mean runtimes of 3--4\,ms, with cold starts below 5\,ms.
The P90 values remain well below the 1\,s acoustic update period, confirming real-time feasibility for the tested scenarios.

\subsection{Scalability Aspects}
\label{sec:scalability_aspects}
We evaluated the whale-like scenario with clean measurements, a 1\,s emission period, and the near initial estimate for different numbers of trackers.
Tracker positions are known inputs, so increasing the number of trackers does not add decision variables: for horizon $N$, the MHE has $5(N+1)$ target-state variables, while the number of scalar TDoA residuals grows as $(M-1)(N+1)$.
Thus, additional trackers mainly increase measurement-residual and Jacobian evaluation.
Table~\ref{tab:exec_time_scalability} reports the timing statistics from three independent runs.

\begin{table}[htbp]
  \centering
  \captionsetup{width=\columnwidth,justification=centering}
  \caption{
    Scalability test for the whale-like scenario.
    Sliding-window statistics report the median and the 90th percentile (P90) over the last $50$ warm-started updates.
  }
  \label{tab:exec_time_scalability}
  \setlength{\tabcolsep}{3pt}
  \renewcommand{\arraystretch}{1.05}
  \begin{tabular}{lccc}
  \toprule
  \multirow{2}{*}{Trackers} 
    & \multirow{2}{*}{Cold start time [ms]} 
    & \multicolumn{2}{c}{Sliding window time [ms]} \\
  \cmidrule(lr){3-4}
    &  & median & P90 \\
  \midrule
  2 trackers   & $3.20$ & $3.88$ & $6.81$ \\
  3 trackers   & $3.74$ & $3.68$ & $5.99$ \\
  5 trackers   & $3.83$ & $3.57$ & $6.57$ \\
  10 trackers  & $4.44$ & $3.82$ & $7.41$ \\
  \bottomrule
  \end{tabular}
\end{table}

According to the simulation data, increasing the number of trackers from $2$ to $10$ raises the residual count from $16$ to $144$, while the median runtime stays within $3.57$--$3.88$\,ms and the P90 remains below $7.5$\,ms.

%% file: part5_conclusion.tex
\section{Conclusion and Future Work}
\label{sec:conclusion}

This paper presented an MHE-based method for underwater target localization and tracking using TDoA measurements.
The approach integrates the nonlinear kinematic model and TDoA structure within a unified optimization framework, enabling consistent state estimation over a receding horizon.
A comparative study with an EKF baseline in a controlled simulation environment captured key characteristics of underwater acoustic tracking, including sparse updates, nonlinear motion, and measurement outliers.

The simulation results show that the proposed MHE estimator maintains stable performance across the tested motion scenarios and is notably more robust than the EKF under poor initialization, sparse measurements, and outliers.
Timing experiments on the tested hardware further indicate real-time feasibility for the considered problem.

Future work includes extending the formulation to three-dimensional motion, accommodating irregular or event-based acoustic measurements, and developing decentralized variants suitable for multi-agent marine robotic systems.